\documentstyle[aps,prd,preprint,floats,epsf,epsfig]{revtex}

\begin{document}

\newcommand{\be}{\begin{equation}}
\newcommand{\ee}{\end{equation}}
\newcommand{\ba}{\begin{eqnarray}}
\newcommand{\ea}{\end{eqnarray}}

\title{Test of factorisation in $B \to K\pi$ decays }
\author{ Claudia Isola and T. N. Pham }
\address{ Centre de Physique Th{\'e}orique, \\Centre National de la Recherche 
Scientifique, UMR 7644,\\ 
Ecole Polytechnique, 91128 Palaiseau Cedex, France\\}

\date{24 November 1999}
\maketitle

\begin{abstract}
We analyse the $B \to K\pi$ decays using the factorisation model with
final state interaction phase shift included. We find that factorisation
seems to describe qualitatively the latest CLEO data. For a test of 
the factorisation model, we derive a relation for the branching
ratios independent of the strength of the strong penguin interactions.
This relation gives a central value of $(0.60 \times 10^{-5})$ for 
${\mathcal B}(\bar{B}^{0} \to \bar{K}^{0}\pi^{0})$, somewhat
smaller than the latest CLEO measurement, but the experimental errors
are yet too big to take it as a  real prediction of the factorisation model.
We also find that a ratio obtained from the 
CP-averaged $B \to K\pi$ decay rates could 
be used
to test the factorisation model and to determine the weak angle $\gamma$
with more precise data, though the latest CLEO data seem to favor
$\gamma$ in the range of $(90^{\circ}-120^{\circ})$.

\end{abstract}

\pacs{PACS numbers: 13.20.He, 11.30.Er, 12.38.Bx}


\narrowtext

One of the possibilities offered by the $B \to K\pi$ decays is the 
determination of the CP-violating phase $\gamma$, one of the angles
in the $(db)$ unitary triangle of the Cabibbo-Kobayashi-Maskawa (CKM)
quark mixing matrix in the standard model \cite{PDG}. Infact the large
CP-averaged
branching ratio(${\mathcal B}$) for  $B \to K\pi$ as observed  by the CLEO
Collaboration \cite{cleo2} indicate that the penguin interactions
contribute a major part to the decay rates and provide an interference
between the Cabibbo-suppressed tree and penguin contribution resulting in
a CP-asymmetry between the $B \to K\pi$ and its charge conjugate
mode. The CP-averaged decay rates  depend also on  the weak phase
$\gamma$ and give us a  determination of this phase once a reliable
description of the $B \to K\pi$ decays could be established
\cite{Fleischer,Neubert}. 

With the latest measurement by the CLEO collaboration \cite{cleo2} , we
have now  the CP-averaged branching ratios for all 
the $B \to K\pi$ decay modes. 
In particular, the $\bar{B}^{0} \to \bar{K}^{0}\pi^{0}$ 
mode is found to have a large branching ratio of 
$(1.46^{+5.9+2.4}_{-5.1-3.3})\times 10^{-5}$  compared with a value
in the range $(0.5 - 0.74) \times 10^{-5}$ in  the factorisation model 
\cite{Deshpande,Ali}. The predicted values for other modes are,
however, more or less in agreement with experiment. As the
effective Hamiltonian for $B \to K\pi$ decays is well e\-sta\-bli\-shed with
the short-distance Wilson coefficients for tree and 
penguin operators now given
at the next-to-leading logarithms(NLL) QCD 
radiative corrections \cite{Ali,Buras,Ciuchini,Fleischer1,Kramer,Deshpande1}, 
the most important
theoretical uncertainties would probably come from 
long-distance matrix elements obtained
with the factorisation model and final state interaction (FSI) effects.
Infact one of the main uncertainties in the penguin contributions  
to $B \to K\pi$ decays come from the value of the current 
$s$ quark mass which is not known to a good accuracy. 
There are also non-factorisation terms
which must be included in the form of an effective Wilson coefficients
to make the amplitudes scale-independent  \cite{Ali,Buras2}. Thus a
more precise test of factorisation is to consider quantities which are
independent of the strong penguin contributions. This is the main
purpose of this paper. When all the $B \to K\pi$  decay modes are
measured with good accuracy, and if the rescattering phase is known 
the dominant strong penguin contribution 
could be determined from the measured branching ratios assuming 
factorisation  for the small tree-level and electroweak penguin terms, as 
will be discussed in the following. Though the present data are not yet
sufficiently accurate for a  determination of
the effective Wilson coefficients  in $B \to K\pi$ decays at this time, 
a first step toward an understanding of  $B \to K\pi$ decays is to see
how well these penguin-dominated charmless $B$ decays  
can be described by factorisation using the Wilson coefficients 
obtained from perturbative QCD.   
As argued in \cite{Neubert2}, for these very 
energetic decays,  because of color transparency, factorisation should be 
a good approximation for $B \to K\pi$ decays if the Wilson coefficients 
are evaluated at a scale $\mu =O(m_{b})$. We 
 could thus proceed to the test of factorisation
bearing in mind that there are possible
scale-dependent corrections from 
non-factorisation terms to be determined with more precise data.
To include FSI effects, as in \cite{Deshpande}, we assume 
that elastic FSI effects can be absorbed
into the two $\Delta I =1/2$  and $\Delta I =3/2$ elastic 
$\pi K \to \pi K$ rescattering phases
$\delta_{1}$ and $\delta_{3}$ taken as free parameters 
and include only inelastic
effects coming from the charm and charmless intermediate state
contributions to the absorptive part of the
decay amplitudes. These inelastic contributions can be included in
the Wilson coefficients of the penguin operators which now have an absorptive  
part and are given in \cite{Fleischer1,Deshpande1,Hou}.  

We begin by first giving predictions in factorisation model
for the $B \to K\pi$ decay
rates and branching ratios in terms of the rescattering phase difference
$\delta$ and for a typical value of the weak phase $\gamma$. As will 
be seen, factorisation seems to produce  sufficient 
$B \to K\pi$ decay rates. We
could thus proceed to a test of the factorization model 
by comparing with experiments, 
quantities obtained by factorisation which are
independent of the strong rescattering phase difference \cite{Kamal}. 
We find that the sum of the CP-averaged
branching ratios 
${\mathcal B}(B^{-} \to K^{-}\pi^{0}) + 
{\mathcal B}(B^{-} \to \bar{K}^{0}\pi^{-})$ 
and ${\mathcal B}(\bar{B}^{0} \to K^{-}\pi^{+}) + 
{\mathcal B}(\bar{B}^{0} \to \bar{K}^{0}\pi^{0})$ are independent of the FSI
rescattering phase. Other quantities obtained from 
various combination of the decay rates, for example, the quantity
 $\Delta $ defined as
$\Gamma(B^{-} \to \bar{K}^{0}\pi^{-}) + \Gamma(\bar{B}^{0} \to K^{-}\pi^{+}) 
-2(\Gamma(B^{-} \to K^{-}\pi^{0}) + \Gamma(\bar{B}^{0} \to
\bar{K}^{0}\pi^{0}))  $  is independent of the strong penguin
contributions and could be used to predict 
${\mathcal B}(\bar{B}^{0} \to \bar{K}^{0}\pi^{0})$ in terms of the other
measured branching ratios. As the main purpose of this paper
is to test the factorisation model using relations
independent of the strong penguin interactions, we will not discuss here
a recent theoretical work on factorisation in $B \to \pi\pi $ decays
which should be completed to have all the 
logarithms of $m_{b}$ under control \cite{Beneke}.   

In the standard model, the effective Hamiltonian for $B \to K\pi$ decays
are given by \cite{Buras,Ciuchini,Deshpande1},
\begin{eqnarray}
H_{\rm eff} &=& {G_F\over \sqrt{2}}[V_{ub}V^*_{us}(c_1O_{1}^{u} + c_2
O_{2}^{u}) + V_{cb}V^*_{cs}(c_1O_{1}^{c} + c_2 O_{2}^{c})  \nonumber\\
&&-\sum_{i=3}^{10}(V_{tb}V^*_{ts} c_i) O_i] 
+ {\rm h.c.}\;,
\label{hw}
\end{eqnarray}
in standard notation.
At next-to-leading logarithms, $c_{i}$ take the form of an effective
Wilson coefficients $c_{i}^{\rm eff}$ which  contain also the penguin 
contribution from the $c$ quark loop and 
are given in \cite{Fleischer1,Deshpande1}.

The tree level operators $ O_{1}$ and $ O_{2}$  as well as the electroweak
penguin operators $O_{7}-O_{10}$ have both $I =0$  and $I=1$ parts while
the QCD strong penguin operators $O_{3}-O_{6}$ have only $I=0$ terms.
The $B \to K\pi$ decay amplitudes can now be expressed in terms of the
decay amplitudes into $I=1/2$ and $I=3/2$ final states as \cite{Deshpande},
\begin{eqnarray}
A_{K^-\pi^0} &=& 
 {2\over 3} B_3e^{i\delta_3} + \sqrt{{1\over 3}} 
(A_1+B_1)e^{i\delta_1}, \ \ \  \nonumber\\
A_{\bar K^0\pi^-} &=&
{\sqrt{2}\over 3} B_3e^{i\delta_3} - \sqrt{{2\over 3}} 
(A_1+B_1)e^{i\delta_1},\nonumber\\
A_{K^-\pi^+} &=&
{\sqrt{2}\over 3} B_3e^{i\delta_3} + \sqrt{{2\over 3}} 
(A_1-B_1)e^{i\delta_1},\ \ \ \nonumber\\
A_{\bar K^0\pi^0} &=& 
{2\over 3} B_3e^{i\delta_3} - \sqrt{{1\over 3}} 
(A_1-B_1)e^{i\delta_1},
\label{amplitude}
\end{eqnarray}
where $A_{1}$ is the sum of the  strong penguin $A_{1}^{\rm S} $
and the $I=0$  tree level $A_{1}^{\rm T}$ as well as the $I=0$ 
electroweak penguin $A_{1}^{\rm W}$ contributions 
to the $B \to K\pi$ $I=1/2$ amplitude;  similarly  
$ B_{1}$ is the sum of the
$I=1$ tree level $B^{T}_{1}$ and electroweak penguin $B^{W}_{1}$
contribution to the $I=1/2$ amplitude, and $B_{3}$ is the sum of the
$I=1$  tree level $B^{T}_{3}$ and electroweak penguin $B^{W}_{3}$
contribution to the $I=3/2$ amplitude. 

The factorisation approximation is obtained by neglecting in the 
Hamiltonian terms which are the product of two
color-octet operators after Fierz reordering of the quark fields. The
effective Hamiltonian for non-leptonic decays are then given by
Eq.(\ref{hw})
with $c_{j}$ replaced by $a_{j}$ and  $O_{j} $ expressed in terms of
hadronic field operators. In the notation of 
Ref.\cite{Deshpande}, we have 
\ba
&&A_1^{\rm T} = i{\sqrt{3}\over 4}\, V_{ub}V_{us}^*\; r\; 
a_2,\nonumber\\
&&B_1^{\rm T} = i {1\over 2\sqrt{3}}\, V_{ub}V_{us}^*\; r 
\left[-{1\over 2}a_2 
+a_1 X\right],\nonumber\\
&&B_3^{\rm T} = i{1\over 2}\, V_{ub}V_{us}^*\; r 
\left[a_2 + a_1 X\right],
\nonumber\\
&&A_1^{\rm S} = -i{\sqrt{3}\over 2}\, V_{tb}V_{ts}^*\; r
\left[ a_4 + a_6 Y\right],
\ \ \ \ \ \ B_1^{\rm S} = B_3^{\rm S} = 0\nonumber\\
&&A_1^{\rm W} = -i{\sqrt{3}\over 8}\, V_{tb}V_{ts}^*\; r
\left[a_8Y +  a_{10}\right],
\nonumber\\
&&B_1^{\rm W} = i{\sqrt{3}\over 4}\, V_{tb}V_{ts}^*\; r
\left[{1\over 2}a_8Y 
+ {1\over 2}a_{10} 
+\left(a_7 - a_9\right)X\right],
\nonumber\\
&&B_3^{\rm W} = - i{3\over 4}\, V_{tb}V_{ts}^*\; r
\left[\left( a_8Y 
+ a_{10} \right) 
-\left(a_7 - a_9\right)X\right],
\ea
where $r = G_F\, f_K F^{B\pi}_0(m^2_K) (m_B^2-m_\pi^2)$,
$X= (f_\pi/f_K)(F^{BK}_0(m^2_\pi)/F^{B\pi}_0(m^2_K)) 
(m_B^2-m^2_K)/(m_B^2-m_\pi^2)$, 
$Y = 2m^2_K/[(m_s+ m_q)(m_b-m_q)]$ with $q=u,\ d$ for 
$\pi^{\pm,0}$ final states, respectively,
 and $a_{j}$ are given in terms of the effective Wilson coefficients
$c^{\rm eff}_{j}$ ($N_{c}$ is the number of effective colors) by
\ba
a_{j} &=& c^{\rm eff}_{j} + c^{\rm eff}_{j+1}/N_{c} \,\, {\rm for} \ j=1,3,5,7,9 \nonumber\\
a_{j} &=& c^{\rm eff}_{j} + c^{\rm eff}_{j-1}/N_{c} \,\, {\rm for} \ j=2,4,6,8,10\,\, .
\ea
In our analysis, we use
$N_c =3$ and $m_{b} = 5.0\; \rm GeV$ which give
$a_{j}$  the following numerical values
\ba
a_1 &=& \ 0.07    , \ \ \ a_2=  \ 1.05  , \nonumber\\
a_4 &=& -0.043 - 0.016i ,\ \ \ a_6= -0.054 - 0.016i  , \nonumber\\
a_7 &=& \ 0.00004 -0.00009i  , \ a_8 =  0.00033 - 0.00003i  , \nonumber\\
a_9 &=& \-0.00907-0.00009i  , \ a_{10}=  -0.0013 -0.00003i .
\label{coeff}
\ea
Note that $a_{1}$ is sensitive to  $N_{c}$ and is rather  small for 
$N_{c} =3$ . As there is no evidence for a large positive $a_{1}$ in
$B \to K\pi$ decays which are penguin-dominated and are not sensitive 
to $a_{1}$, we use $a_{1}$ evaluated with $N_{c} =3$ given in 
Eq.(\ref{coeff}). Indeed, the predicted branching ratios remain
essentially unchanged with $a_{1}=0.20$ taken from the Cabibbo-favored 
$B$ decays \cite{Kamal1,Neubert1}. 

In the absence of  FSI rescattering phases, we recover the usual
expressions for the decay amplitudes in the factorisation approximation.
We have used   $c^{\rm eff}_{j}$ given at next-to-leading order in 
QCD radiative corrections \cite{Buras,Ciuchini,Deshpande1} and 
evaluated at a scale $\mu = m_{b}$. We note that the coefficients 
$c^{\rm eff}_{3}$,
$c^{\rm eff}_{4}$, $c^{\rm eff}_{5}$ and $c^{\rm eff}_{6}$ 
are enhanced by the internal charm quark loop  due to the large
time-like virtual gluon momentum $q^{2} = m_{b}^{2}/2$ as pointed out
in \cite{Fleischer,Fleischer1,Hou} (the other electroweak penguin 
coefficients like $c^{\rm eff}_{7}$  and $c^{\rm eff}_{9}$ are not
affected by this charm quark loop contribution in any significant amount).
This enhancement of the strong penguin term increases the decay rates and 
bring the theoretical $B \to K\pi$ decay rates closer to the latest
CLEO measurements. 
In the above expressions, the tree level amplitudes are suppressed
relative to the penguin terms by the CKM factor 
$V_{ub}V_{us}^*/V_{tb}V_{ts}^* $ which can be approximated by
$-(|V_{ub}|/|V_{cb}|)\times (|V_{cd}|/|V_{ud}|)\exp(-i\;\gamma)$
after neglecting terms of the order $O(\lambda^{5})$ in the (bs)
unitarity triangle.  
The $B \to K\pi$ decay rates then depend on the FSI rescattering
phase difference $\delta = \delta_{3} -\delta_{1}$ and the weak phase
$\gamma$ . In the following, 
we shall use the set of parameters
 of \cite{Ali,Bauer} which give
$f_{\pi} =133 \;\rm MeV$, $f_{K} =158\;\rm MeV$, 
$ F_{0}^{B\pi}(0) = 0.33$, $ F_{0}^{BK}(0) = 0.38$. We use  $m_{s} =
120\rm MeV$, $|V_{cb}|=0.0395, |V_{cd}|=0.224 $
and $|V_{ub}|/|V_{cb}|=0.08$ \cite{PDG}. At the moment, $m_{s}$ is not
known to a good accuracy, but a value around $100-120\,\rm MeV$
inferred from $m_{K^{*}} - m_{\rho}$, $m_{D_{s}^{+}}- m_{D^{+}}$ 
and $m_{B_{s}^{0}}- m_{B^{0}}$ 
mass differences
\cite{Colangelo} seems not unreasonable. 
To show  the factorisation predictions and the  dependence
of the branching ratios on the rescattering phase difference
$\delta $, we give, as an example, the CP-averaged  
$B \to K\pi$ decay rates in Fig.1 evaluated
with a  CKM value given by $\rho = 0.12$, $\eta =0.34 $ \cite{Ali}
corresponding to  $\gamma = 70^{\circ} $.

As can be seen from Fig.1,
all the $B \to K\pi $ decay modes for $B^{-}$ and $\bar{B}^{0}$ ,
except the $\bar{B}^{0} \to \bar{K}^{0}\pi^{0}$ 
mode, have branching ratios more or less in agreement with 
the latest CLEO data \cite{cleo1,cleo2} which give, for the CP-averaged 
branching ratios
\begin{eqnarray}  
{\mathcal B} (B^{+} \rightarrow K^{+} \pi^0)
&=&(11.6^{+3.0+1.4}_{-2.7-1.3}) \times 10^{-6},  \nonumber  \\
{\mathcal B} (B^{+} \rightarrow K^0 \pi^{+})  &=& 
(18.2^{+4.6}_{-4.0}\pm 1.6) \times 10^{-6},   \nonumber  \\
{\mathcal B} (B^0\rightarrow K^{+} \pi^{-})  &=& 
(17.2^{+2.5}_{-2.4}\pm 1.2) \times 10^{-6}, \nonumber \\ 
{\mathcal B} (B^0 \rightarrow K^0 \pi^0)  &=&
(14.6^{+5.9+2.4}_{-5.1-3.3}) \times 10^{-6}.
\end{eqnarray}

The computed decay rates shown above could be larger with the form
factors given
in \cite{Casalbuoni} and could bring the 
$B \to K\pi $ decay rates  closer to the latest CLEO data.

We now turn to the test of factorisation  in $B \to K\pi $
decays. The decay rates into a $K\pi$ final state is given by
\be
\Gamma(B \to K\pi) = C|A_{K\pi}|^2
\ee
where the subscript  $K\pi $ refers to any of the decay modes for $B^{-}$
and $\bar{B}^{0} $ and $C$ is the usual phase space factor. By summing over
the decay modes for $B^{-} $ and for $\bar{B}^{0} $ 
respectively, we have in terms of $A_{1} $, $B_{1} $ and $B_{3} $ ,
\ba
\Gamma_{B^-} & = & C\left[{2\over 3}|B_3|^2 +|A_1+B_1|^2\right]\nonumber\\   
\Gamma_{B^0} & = & C\left[{2\over 3}|B_3|^2 +|A_1-B_1|^2\right]\, .
\label{totalrates}
\ea
where $\Gamma_{B^-} = \Gamma(B^{-} \to K^- {\pi}^0) + 
\Gamma(B^{-} \to \bar{K}^{0} {\pi}^{-})$ and 
$\Gamma_{B^0} = \Gamma(\bar{B}^{0} \to K^- {\pi}^+) + 
\Gamma(\bar{B}^{0} \to \bar{K}^{0} {\pi}^{0})$.

The quantities $\Gamma_{B^-} $ and $\Gamma_{B^0} $ are independent of
the rescattering phase difference $\delta $. They are given in the
factorisation model as a function of the weak phase $\gamma $. Two other
related quantities of interest obtained from the above Eq.(\ref{totalrates}) 
are 
\ba
r_b\,{\cal B}_{B^-} + {\cal B}_{B^0} & = & 2\,C\left[{2\over 3}|B_3|^2 +|A_1|^{2}
+|B_1|^2\right]\tau_{B^{0}}\nonumber\\
r_b\,{\cal B}_{B^-} - {\cal B}_{B^0} & = & 4\,C\,{\rm Re}\,(A_1^{*}B_1)\tau_{B^{0}}\, 
\label{ttrates}
\ea

which, together with one measured $B \to K\pi $ branching ratio, would
enable us to determine the strength of the
strong penguin contribution as well as its absorptive part and $\gamma$
assuming factorisation for 
the small tree-level and electroweak penguin contributions, if the
rescattering phase difference $\delta$ could be inferred from the 
$\delta$-dependent branching ratio and from other sources. In the above
expression, $\tau_{B^{0}} $ is the $ B^{0}$ lifetime and 
$r_b=\tau_{B^0}/ \tau_{B^-}$,
 
Also, if all the four $B \to K\pi$ decay rates (CP-averaged) are 
measured with good accuracy, in particular with a precise measurement
of the $\bar{B}^{0} \to \bar{K}^{0} {\pi}^{0} $ branching ratio, 
the following quantities
\be
R_{1} = {\Gamma_{B^{-}} \over \Gamma_{B^0} } \, \ ,\
R_{2} = {\Gamma_{B^{-}} \over (\Gamma_{B^{-}} +\Gamma_{B^0})}
\label{ratios}
\ee
could also be used to test factorisation. 

As the CP-averaged $B \to K\pi$ decay rates depend  on $\gamma$,
the  computed partial rates $\Gamma_{B^{-}} $ 
and $\Gamma_{B^0} $ would now lie
between the upper and lower limit corresponding to $\cos(\gamma) =1 $ 
and $\cos(\gamma) =-1 $, respectively.
As shown in Fig.2, where the corresponding CP-averaged branching
ratios (${\mathcal B}_{B^0} $ and
${\mathcal B}_{B^-}$)
for $\Gamma_{B^{-}} $ and $\Gamma_{\bar{B}^{0}} $ are plotted against 
$\gamma$, the factorisation model values with the 
BWS form factors \cite{Bauer}
seem somewhat smaller
than the CLEO central values by about $10-20\% $.  
Also,  ${\mathcal B}_{B^{-}} > {\mathcal B}_{\bar{B}^{0}} $
while the data gives 
${\mathcal B}_{B^{-}} < {\mathcal B}_{\bar{B}^{0}} $
by a small amount which could be due to a large
measured $\bar{B}^{0} \to \bar{K}^{0}\pi^{0} $ decay rates. 
We note that smaller values for the form factors 
could easily accommodate the latest CLEO measured values, if a smaller
value for $m_{s}$, e.g, in the range $(80-100)\,\rm MeV$ is used. What
one learns from this analysis is that $B \to K\pi$ decays are 
penguin-dominated and the strength of the penguin interactions as obtained by
perturbative QCD, produce sufficient $B \to K\pi$ decay rates and that
factorisation seems to work with an accuracy better than a factor of 2,
considering large uncertainties from the form factors and possible
non-factorisation terms inherent in the factorisation model. With more
precise measuremnts expected in the near future, it might be possible to
have a detailed test of factorisation and a determination of  $\delta$    
and $\gamma$ by comparing with experiments various
relative branching ratios, to reduce  
uncertainties from  form factors and CKM parameters. 
Other test of factorisation could also be done by looking for quantities
which  are independent of the strong penguin interactions. Infact, since
the four $B \to K\pi$ decay rates depend on only three amplitudes
$A_{1} $, $B_{1} $ and $B_{3} $, it is possible to derive a relation
between the decay rates independent of $A_{1}$. From the following quantities,

\ba
\lefteqn{\Gamma(B^{-} \to \bar{K}^{0} {\pi}^{-}) +\Gamma(\bar{B}^{0}
\to K^- {\pi}^+)=C_1} \nonumber \\ 
&&\times  \left[{1\over 3} |B_3|^2 +(|A_1|^2 +
|B_1|^2)-{2\over \sqrt 3}{\rm Re}(B_3^* B_1 {e}^{i\delta})\right]\nonumber\\
\lefteqn{\Gamma(B^{-} \to K^- {\pi}^0)+\Gamma(\bar{B}^{0} \to
     \bar{K}^{0} {\pi}^{0}) =C_2} \nonumber\\
&&\times \left[{4\over 3} |B_3|^2 +(|A_1|^2 +
|B_1|^2)+{4\over \sqrt 3}{\rm Re}(B_3^* B_1 {e}^{i\delta})\right]
\ea
where $C_1 ={4\over 3}C$ and $C_2 ={2\over 3}C$, we obtain
\ba
\Delta & = &\left\{ \Gamma(B^{-} \to \bar{K}^{0} {\pi}^{-})+\Gamma(\bar{B}^{0}
\to K^- {\pi}^+) \right. \nonumber\\
   & - & \left. 2\left[\Gamma(B^{-} \to K^- {\pi}^0)+\Gamma(\bar{B}^{0} \to
     \bar{K}^{0} {\pi}^{0})\right]\right\}\tau_{B^0}\nonumber\\
& = & \left[-{4\over 3}|B_3|^2-{8\over {\sqrt 3}}{\rm Re} (B_3^* B_1 {\rm
  e}^{i\delta})\right](C\tau_{B^0})\,\, 
\label{delta}
\ea

From Eq.(\ref{delta}), we see that $\Delta$  is independent of $A_1$ and hence
is independent of the strong penguin term. It is given by the
tree-level and electroweak  contributions which are much smaller
than the strong penguin term. As can be seen 
from Fig.2, where its values for  $\delta =0$ 
are plotted against $\gamma $. $\Delta$ is of the order $O(10^{-6})$
compared with  ${\mathcal B}_{B^{-}} $ and ${\mathcal
  B}_{\bar{B}^{0}} $ which are dominated by the strong penguin
contribution and are in the range  $(1.6-3.0)\times 10^{-5}$. As the variation with $\delta$ is negligible, $\Delta$ remains at the 
$O(10^{-6})$ level for other values of $\delta \not= 0$. Thus, to this
level of accuracy, we can put $\Delta \simeq 0$. Eq.(\ref{delta}) becomes
\be
{ r_b}{\mathcal B}_{\bar{K}^{0}
    {\pi}^{-}}+{\mathcal B}_{K^- {\pi}^+} = 
2\,\left[{\mathcal B}_{\bar{K}^{0} {\pi}^{0}}+
{ r_b}{\mathcal B}_{ K^- {\pi}^0}\right]\,.
\label{kpi0}
\ee
This relation can be used as
a test of factorization with more 
precise measurements of the CP-averaged
branching ratios. Conversely, it can also be used to predict 
 ${\mathcal B}(\bar B^0\to \bar K^0 {\pi}^0)$ in terms of the 
other measured branching ratios.
From the latest CLEO data, with $\Delta \simeq 0 $,
Eq.(\ref{kpi0}) then gives     
${\mathcal B}(\bar{B}^{0} \to\bar{K}^{0}
{\pi}^{0})= 0.60\times 10^{-5}$. As can be seen, the
large experimental errors prevent us from drawing any firm conclusion
on the validity of factorisarion, though, the above predicted central value
for ${\mathcal B}(\bar B^0\to \bar K^0 {\pi}^0)$ is somewhat smaller
than the CLEO data.  

For another test of factorisation and a 
determination of $\gamma$
we have derived a relation in the form of a ratio which is 
independent of the form factors and the CKM parameters. It is given by 
the ratio $R$ of the two CP-averaged quantities as
\be
R={{\left[{\mathcal B}(B^-\to K^{-} {\pi}^{0})
+{\mathcal B}( B^-\to \bar{K}^{0} {\pi}^{-})\right]}\over{
{\mathcal B}(B^-\to\bar{K}^{0}
    {\pi}^{-})+{{\mathcal B}(\bar B^0\to K^- {\pi}^+)}/{r_b}}}\,\, .
\label{R}
\ee
Numerically, we find that terms  proportional to $\cos(\delta)$ and 
$\sin(\delta)$ in $R $ is of the order $10^{-7}$ and thus can be safely
ignored. Thus $R $ is a function of $\gamma$ alone and can be used to
determine $\gamma $ as it does not suffer
from the uncertainties in the form factors and in the CKM parameters. 
In Fig.3 we give a plot of $R $ as a function of $\gamma$. As can be
seen, it is not possible to deduce a value for $\gamma$ with the 
CLEO data which gives $R=(0.80\pm 0.25) $ 
as the theoretical prediction for $R$ lies within
the experimental errors. If we could reduce the experimental
uncertainties to a level of less than $10\%$, we might be able to 
give a value for $\gamma$. Thus it is important to measure 
$B\to K\pi$ decays branching ratios to a high precision.  
It is interesting to note that the central
value of $0.80$ for $R$ corresponds to $\gamma = 110^{\circ}$, close to 
the   value  $(113^{+25}_{-23})^{\circ}$ found
by the CLEO Collaboration in an analysis of all known charmless two-body
$B$ decays with the factorisation model \cite{cleo1}. It seems that 
the CLEO data favors a large  
$\gamma$ in the range $(90^{\circ}-120^{\circ})$. A large
$\gamma$ as  shown in Fig.2, would increase the factorisation values 
for ${\mathcal B}_{B^-} $ and ${\mathcal B}_{B^0}$ which are given
numerically by 
\ba
&&{\mathcal B}_{B^-} = (2.757 - 0.409\cos(\gamma)) \times 10^{-5}\ ,\nonumber\\
&&{\mathcal B}_{B^0}= (2.270 - 0.624\cos(\gamma))\times 10^{-5} \,\ .
\label{rates1}
\ea
For the ratio $R$, we have
\be
R = {(2.651 - 0.393\cos(\gamma)) \over (3.253 - 0.652\cos(\gamma))}
\ee
Also shown in Fig.3 are the ratio $R_{1}$ and $R_{2}$ defined in 
Eq.(\ref{ratios}). As $R_{1}$ shows strong dependence on $\gamma$, a
better way to determine $\gamma$ would be to use $R_{1}$ rather than
$R$ when a precise value for ${\mathcal B}(\bar B^0\to \bar K^0 {\pi}^0)$
will be available.

Given $\gamma = 110^{\circ}$, all the $B \to K\pi$ branching ratios
can be predicted in terms of the rescattering phase difference $\delta$
as shown in Fig.4. Comparing with Fig.1, we see that, except for
${\mathcal B}(\bar B^0\to \bar K^0 {\pi}^0)$ which remains at the 
$6.5\times 10^{-6}$ level, the other branching ratios become larger
with $\gamma = 110^{\circ}$ 
and closer to the CLEO data which indicate $B^-\to \bar K^0 \pi^-$
and $\bar B^0 \to K^-\pi^+$ are the two largest modes with near-equal
branching ratios in qualitative agreement with factorisation. 
Fig.1 shows that these two largest branching ratios are quite apart,
except for $\delta < 50^{\circ}$ while Fig.4 suggests 
$\delta $ should be large, in the range of $(40^{\circ}-70^{\circ})$.
With a smaller $\gamma < 110^{\circ}$ and some adjustment of form 
factors, the current $s$ quark mass and 
CKM parameters, it might be possible to accommodate these two largest
branching ratios with a smaller $\delta $. We note that
the dependence of the four branching ratios shown in Fig.1 
and Fig.4 are essentially the same and is given by
$\rm (4/3\sqrt{3})\,Re(A_{1}B_{3}^{*}\exp(i\delta))$,  apart
from the sign, as the interference term 
$\rm Re(B_{1}B_{3}^{*}\exp(i\delta)) $ 
is much smaller than $\rm Re(A_{1}B_{3}^{*}\exp(i\delta))$.

We note that we have also considered  a possible contribution from
inelastic rescattering effects as an additional small absorptive contribution
$A_{i}$ to $A_1$  from  $D\;D^{*}_{s}$ and other
intermediate states to  the $S$-matrix 
unitarity relation. We find that the variation of
$\Gamma_{B^-}$ and $\Gamma_{B^0}$ as a function of $A_{i}$ is 
 negligible. For this reason,
we have set $A_{i}=0$. Also, since the theoretical values 
for the decay rates shown
above show qualitative agreement with the measured values, the strong
penguin terms with enhancement by the internal $c$-quark loop seem to
produce sufficient decay rates, a large dispersive inelastic contribution 
would not be needed in  $B \to K\pi $ decays.

The CP-asymmetries, plotted against
$\delta$ as shown in Fig.5, are given by
\be
 As_{\rm CP} = \frac{ \Gamma - \bar{\Gamma}}
{\Gamma + \bar{\Gamma}}
\label{asym}
\ee     
where $\Gamma$ is the decay rate.
The predicted 
CP-asymmetry $As_{\rm CP}$ for the $B \to K\pi $ decay 
modes are in the range $\pm(0.04)$ to $\pm(0.3)$
for the preferred values of $\delta$ in the range
$(40-70)^{\circ}$ mentioned above. The 
latest CLEO measurements \cite{cleo3} however, do not show any 
large CP-asymmetry in $B \to K\pi$ decays, but the errors are still
too large to draw any conclusion at the moment. 
\bigskip

In conclusion, factorisation with enhancement of the strong penguin
contribution  seems to describe qualitatively
the $B \to K\pi$ decays, although the predicted  
${\mathcal B}(\bar B^0\to \bar K^0 {\pi}^0)$ is below the measured value.
Further measurements will enable
us to have a more precise test of factorisation and a determination of
the weak angle $\gamma $ from the FSI phase-independent relations we
shown above.

\vspace{0.5cm}
{\bf Acknowledgments}

We would like to thank Dr Kwei-Chou Yang for pointing out an error in
our initial numerical calculations.

\newpage
\begin{figure}[hbp]
\centering
\leavevmode
\epsfxsize=1.5in
\epsffile{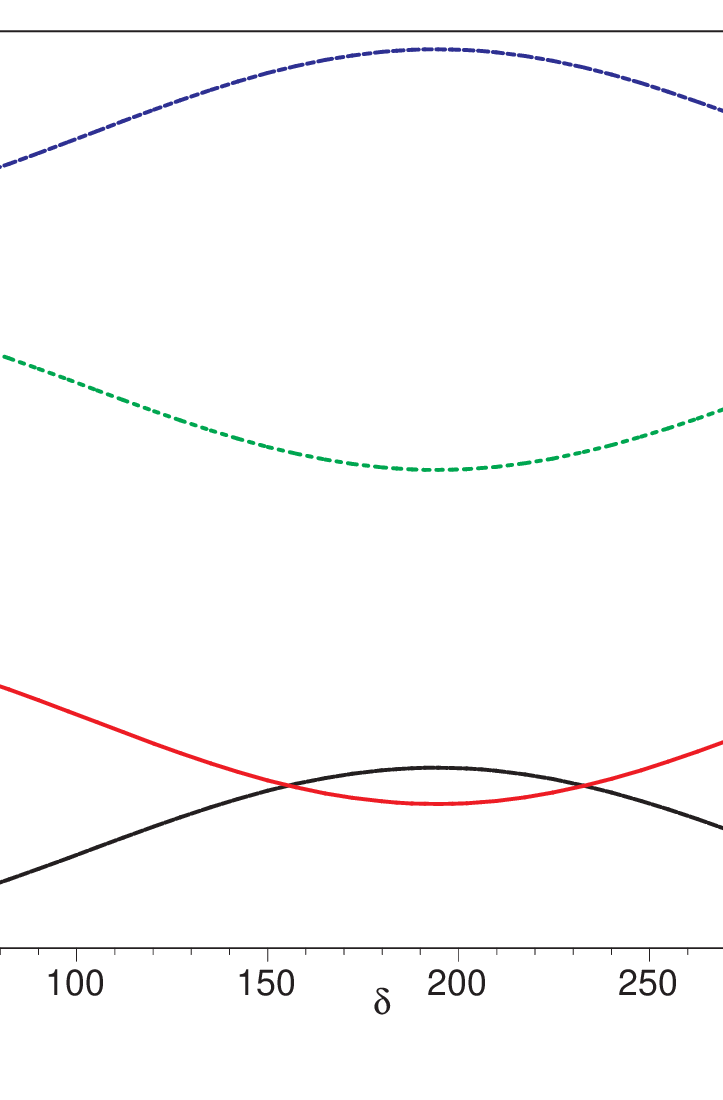}
\caption{${\mathcal B}(B\to K\pi)$ vs. $\delta$ for $\gamma = 70^\circ$.
The curves  
(a), (b), (c), (d)  are for the CP-averaged branching ratios
$B^-\to K^-\pi^0,\ \bar K^0 \pi^-$ and 
$\bar B^0 \to K^-\pi^+,\ \bar K^0\pi^0$, respectively.}
\label{fig:BR}
\end{figure}
\begin{figure}[hbp]
\centering
\leavevmode
\epsfxsize=1.5in
\epsffile{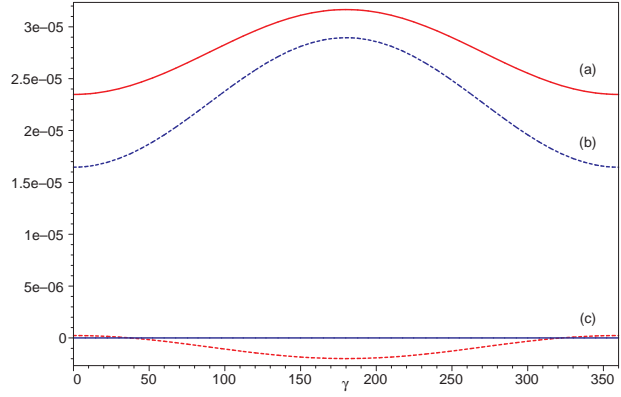}
\caption{${\mathcal B}_{B^-}$ (a), ${\mathcal B}_{\bar{B}^0}$ (b), $\Delta$ (c) vs. $\gamma$}
\label{fig:facttest}
\end{figure}
\begin{figure}[hbp]
\centering
\leavevmode
\epsfxsize=1.5in
\epsffile{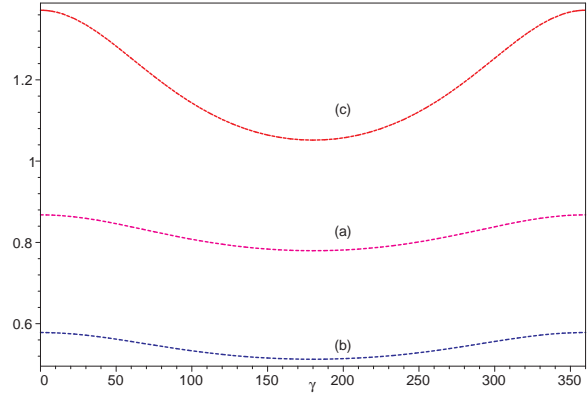}
\caption{The curves (a), (b), (c) are for $R$, $R2$, $R1$ respsectively}
\label{fig:R}
\end{figure}

\begin{figure}[hbp]
\centering
\leavevmode
\epsfxsize=1.5in
\epsffile{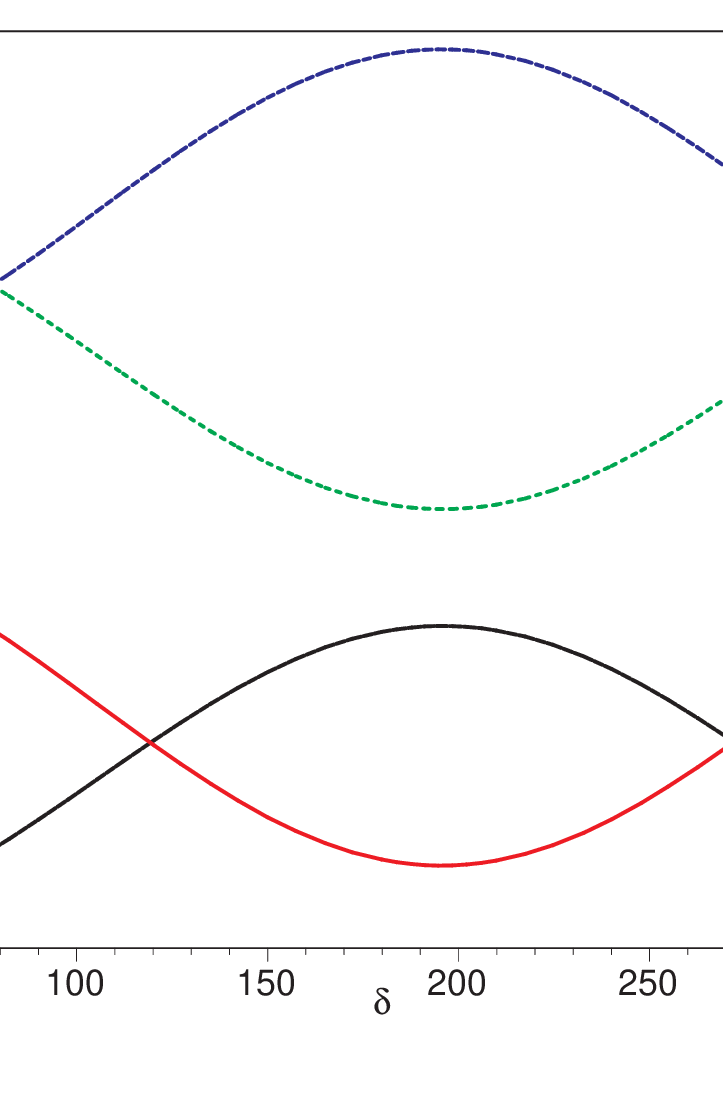}
\caption{${\mathcal B}(B\to K\pi)$ vs. $\delta$ for $\gamma = 110^\circ$.
The curves  
(a), (b), (c), (d)  are for the CP-averaged branching ratios
$B^-\to K^-\pi^0,\ \bar K^0 \pi^-$ and 
$\bar B^0 \to K^-\pi^+,\ \bar K^0\pi^0$, respectively.}
\label{fig:BRc}
\end{figure}
\begin{figure}[hbp]
\centering
\leavevmode
\epsfxsize=1.5in
\epsffile{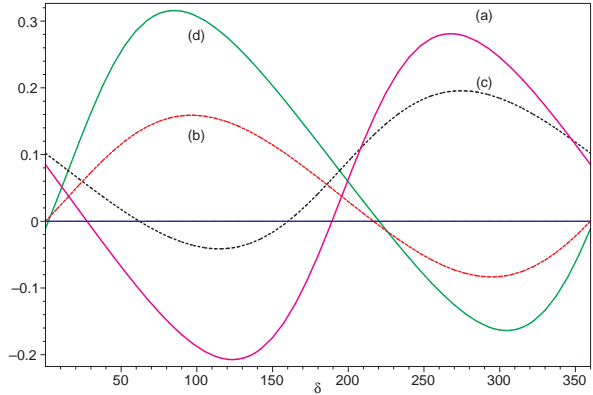}
\caption{The asymmetries vs.$\delta$ for $\gamma = 110^\circ$. 
The curves 
(a), (b), (c), (d)  are for $As_{B^-\to K^-\pi^0}$,$As_{B^-\to\bar
 K^0 \pi^-}$, $As_{\bar B^0 \to K^-\pi^+}$, $As_{\bar B^0 \to \bar K^0\pi^0}$,}
\label{fig:asymm}
\end{figure}

\end{document}